\documentclass[prd,aps,twocolumn,superscriptaddress,preprintnumbers,nofootinbib,showpacs]{revtex4}
\usepackage{graphicx}
\usepackage{exscale}
\usepackage[intlimits]{amsmath}
\usepackage{amsfonts}
\usepackage{amssymb,amscd}
\usepackage{epsfig}
\usepackage{pstricks}

\newcounter{commentdepth}
\newcommand{\be}{\begin{equation}}
\newcommand{\ee}{\end{equation}}

\newcommand{\beqa}{\begin{eqnarray}}
\newcommand{\eeqa}{\end{eqnarray}}
\newcommand{\beq}{\begin{equation}}
\newcommand{\eeq}{\end{equation}}

\newcommand{\p}{\partial}

\newcommand{\reftitle}[1]{}

\begin{document}

\title{Some exact properties of the gluon propagator}


\author{Daniel~Zwanziger}
\affiliation{Physics Department, New York University, New York, NY 10003, USA}

\begin{abstract}
\noindent

Recent numerical studies of the gluon propagator in the minimal Landau and Coulomb gauges in space-time dimension 2, 3, and 4 pose a challenge to the Gribov confinement scenario.  In these gauges all configurations are transverse, $\p \cdot A = 0$, and lie inside the Gribov region $\Omega$, where the Faddeev-Popov operator, $M(A) = - \p_\mu D_\mu(A)$, is positive, that is, $(\psi, M(A) \psi) \geq 0$ for all $\psi$.

We prove, without approximation, that for these gauges, the continuum gluon propagator $D(k)$ in $SU(N)$ gauge theory satisfies the bound ${d-1 \over d} {1 \over (2 \pi )^d} \int d^dk \ {D(k) \over k^2 } \leq N$.  This holds for Landau gauge, in which case $d$ is the dimension of space-time, and for Coulomb gauge, in which case $d$ is the dimension of ordinary space and $D(k)$ is the instantaneous spatial gluon propagator.  This bound implies that $\lim_{k \to 0}k^{d-2} D(k) = 0$, where $D(k)$ is the gluon propagator at momentum $k$, and consequently $D(0) = 0$ in Landau gauge in space-time $d = 2$, and in Coulomb gauge in space dimension $d = 2$, but $D(0)$ may be finite in higher dimension.  These results are compatible with numerical studies of the Landau-and Coulomb-gauge propagator.

In 4-dimensional space-time a regularization is required, and we also prove an analogous bound on the lattice gluon propagator, ${1 \over d (2 \pi)^d} \int_{- \pi}^{\pi} d^dk  { \sum_\mu \cos^2(k_\mu/2) \ D_{\mu \mu}(k)  \over 4 \sum_\lambda \sin^2(k_\lambda/2) } \leq N$.  Here we have taken the infinite-volume limit of lattice gauge theory at fixed lattice spacing, and the lattice momentum componant $k_\mu$ is a continuous angle $- \pi \leq k_\mu \leq \pi$.  Unexpectedly, this implies a bound on the {\it high-momentum} behavior of the continuum propagator in minimum Landau and Coulomb gauge in 4 space-time dimensions which, moreover, is compatible with the perturbative renormalization group when the theory is asymptotically free.  \\
\\
\end{abstract}


\maketitle

\section{Introduction}

	The successes of perturbative calculations at high energy and of numerical studies in lattice gauge theory provide strong evidence that the interactions of quarks and gluons are correctly described by the non-Abelian gauge theory known as QCD.  However we lack a satisfactory understanding of the mechanism, by  which quarks and gluons are confined, comparable to that provided by the Higgs model of electro-weak interactions.  There are several suggestive scenarios which involve the dual Meissner effect with condensation of magnetic monopoles, the maximal Abelian gauge, the maximal center gauge or the light-cone gauge.  
	
	There is also a scenario in Landau and Coulomb gauges that originated with Gribov \cite{Gribov:1977wm} that is based on the insight that there exist Gribov copies --- that is to say gauge-equivalent configurations that satisfy the gauge condition --- and moreover that the dynamics is strongly affected if one cuts off the integral over configurations $A$ to avoid over-counting these copies.  According to this scenario, the cut-off is nearby in infrared directions (in $A$-space), which  suppresses the gluon propagator $D(k)$ at small $k$, so that would-be massless gluons exit the physical spectrum and are said to be confined.  However recent numerical studies, which we shall review shortly, have revealed that the behavior of the gluon propagator is more complicated than expected, and present a challenge to this scenario.  In the present article we present exact bounds on the gluon propagator, which result from the cut-off in $A$-space, that are consistent with and clarify the results found in numerical studies of the gluon propagator.   	
	
    There had been various early  conjectures about the behavior of the gluon propagator $D(k)$.  Gribov in particular obtained by an approximate calculation in Landau gauge \cite{Gribov:1977wm} $D(k) = {k^2 \over k^4 + \gamma}$ in all space-time dimensions.  This has the notable property that $D(0) = 0$, in striking contrast to the tree-level gluon propagator, $D(k) = {1 \over k^2}$, which has a pole at $k = 0$.  This same propagator is also the zeroth-order gluon propagator in a perturbative expansion based on a local, renormalizable action that includes a cut-off at the Gribov horizon \cite{Zwanziger:1989mf}.  However according to numerical studies in Landau gauge it appears that $D(0)$ does not vanish in dimension 2+1 and 3+1.  To address this problem, a dynamically refined action for dimension 2+1 and~3+1 has been proposed and studied \cite{Dudal:2007cw,Dudal:2008rm,Dudal:2008sp,Dudal:2008xd,Dudal:2011gd}.

The gluon propagator has been much studied by Dyson-Schwinger equations and related methods.  The general consensus at present is that there are two types of solution, a scaling solution, with $D(0) = 0$, and a decoupling solution, with $D(0) > 0$, as discussed in \cite{Fischer:2008uz}, where further references may be found.

The gluon propagator in Landau gauge has also been the subject of numerical study in lattice gauge theory.  From recent studies, a somewhat puzzling picture has emerged.  It appears that in dimension 1+1 the gluon propagator in Landau gauge does in fact vanish at $k = 0$, $D(0) = 0$, in accordance with Gribov's  result \cite{Maas:2007uv,Cucchieri:2007rg,Cucchieri:2011um}.  In dimension 2+1 it was found that the gluon propagator has a turn-over below which $D(k)$ {\em decreases} with decreasing $k$,  and there is no explanation for this non-intuitive behavior besides the proximity of the Gribov horizon in infrared directions.  However from studies on huge lattices, it appears that in dimension 2+1, $D(k)$ approaches a {\em finite} value at $k = 0$ \cite{Cucchieri:2007md, Cucchieri:2007rg,Cucchieri:2010xr}.  In dimension 3+1 there may be a kind of shoulder at low momentum and, as in dimension 2+1, it appears that $D(0)$ is finite \cite{Cucchieri:2007md,Bogolubsky:2007ud,Cucchieri:2007rg,Bogolubsky:2009dc,Bornyakov:2009ug}.

In Coulomb gauge the numerical results for the equal-time space-space propagator $D({\bf k})$ in space-time dimension $d+1$  are qualitatively similar to those in Landau gauge in space-time dimension $d$, and infrared suppression is more pronounced for $d = 2$ than for $d = 3$, with  a more distinct decrease with $|{\bf k}|$ at small $\bf k$ for $d = 2$ \cite{Cucchieri:2007cu, Burgio:2008, Burgio:2009, Nakagawa:2011, Maas:2011}.\footnote{Reference \cite{Maas:2011} also presents results for gauges that interpolate between Coulomb and Landau gauges, with gauge condition $\lambda \p_0 A_0 + \p_i A_i = 0$.}  Most of these numerical studies of Coulomb gauge are for $SU(2)$ gauge theory in dimension 3+1, with results for dimension 2+1 in \cite{Burgio:2009, Maas:2011} and for $SU(3)$ gauge theory in dimension 3+1 in \cite{Nakagawa:2011}.  However from numerical studies there is as yet no definite conclusion as to whether in Coulomb gauge $D(0)$ vanishes in dimension 2+1 and 3+1.

To summarize, we are faced with the puzzle that numerical studies in Landau gauge indicate that $D(0) = 0$ in dimension 1+1, which accords with the Gribov scenario, but $D(0)$ is positive, $D(0) > 0$, in dimension 2+1 and 3+1.  The puzzle deepens in view of an argument that led to the conclusion that $D(0) = 0$ in any number of dimensions \cite{Zwanziger:1991}.  This argument involves the free energy, $W(J) \equiv \ln \langle e^{(J, A)} \rangle$, where $J$ is an external source.  The input for this argument is (i) a bound on $W(J)$, and (ii) the hypothesis that $W(J)$ is analytic in $J$.  The bound on~$W(J)$ appears unassailable, and so, if the numerical results are accepted, it must be that $W(J)$ is non-analytic in $J$, which is the signal for a change of phase.  The non-analyticity of $W(J)$ will be discussed elsewhere \cite{Maas:2012}.  

    In the present article we take up another bound, which is found in Appendix B of \cite{Zwanziger:1991}, that does not involve the free energy $W(J)$.  It is an ellipsoidal bound, satisfied by all configurations $A$ in the Gribov region $\Omega$.  From this bound on configurations $A$, we will obtain a bound on the gluon propagator $D(k)$ by taking expectation values.  
    
    In sect.\ IIA we recall some known results, in sect.\ IIB we explain the simple idea on which the derivation of the bound is based, and in sect.\ IIC we note that the bound applies to other gauge bosons.  In sect.\ III we present exact bounds on the gluon propagator that hold in continuum and in lattice gauge theory, and we also present the renormalized continuum bound that holds in dimension 3+1.  In sect.\ IV we discuss the implications of these bounds for the Landau- and Coulomb-gauge propagator in dimension 1+1, 2+1 and 3+1.  In addition to infrared bounds that are stronger in lower space-time dimension, we shall find, unexpectedly, that the cut-off at the Gribov horizon implies a bound on the {\it high}-momentum behavior of the Landau- and Colulomb-gauge propagator in dimension 3+1 given in eqs.~\eqref{UVbond} and \eqref{UVcoulomb} respectively.  Some concluding remarks may be found in sect.\ V.  In Appendix A we derive an ellipsoidal bound on continuum configurations that lie inside the Gribov horizon.  In Appendix B we convert this into a bound on the continuum gluon propagator.\footnote{The same continuum bound is also obtained in eq.~\eqref{regularizedboundd} as a limit of the lattice bound derived in Appendix D, but we have provided an independent derivation of the continuum bound because it is simpler.}  In Appendix C we exhibit a simpler ellipsoidal bound on continuum configurations in the infinite-volume limit.  In Appendix D we derive the bound on the lattice gluon propagator.

\section{Set-up and basic idea}

\subsection{Elementary properties}

	We deal with Euclidean QCD in its continuum and lattice formulations.  Numerical gauge fixing on the lattice is done by gauge transforming to a local minimum of a lattice analog (specified in Appendix D) of the continuum minimizing functional~\cite{Nakajima:1978, Semenov:1982, Zwanziger:1982} 
\beqa
F_A(g) & = & || {^g}A||^2
\nonumber  \\
 & = & \int d^d x \ |{^g}A(x)|^2.
\eeqa
This is the Hilbert square norm of the configuration ${^g}A_\mu = g^{-1} A_\mu g + g^{-1} \p_\mu g$, which is the gauge-transform of the gauge field $A_\mu(x) = t^b A_\mu^b(x)$ by the local gauge transformation $g(x) \in SU(N)$, where $\mu = 1, ... d$.  Here the $t^a$ are an anti-hermitian basis, $(t^a)^\dag = - t^a$, of the Lie algebra of the $SU(N)$ group, $[t^a, t^b] = f^{abc} t^c$, normalized to ${\rm tr}(t^a t^b) = - \delta^{ab}/2$, where $a = 1, ... \ N^2 -1$.  If $d$ is taken to be the dimension of Euclidean space-time, then this gauge fixing produces a gauge in the class of minimal Landau gauges, whereas if $d$ is the dimension of ordinary space, then the gauge is in the class of minimal Coulomb gauges.  (For the Coulomb gauge this minimization is done at every Euclidean time $t$.)  The minimization produces a local minimum of the minimizing functional.  Any local minimum will do.  In principle it could be the absolute minimum, but this is not necessary for our purposes, nor is it achievable in practice numerically.

  At a local minimum (i) the functional $F_A(g)$ is stationary, and (ii) the matrix of its second derivatives is positive.  Property (i) gives the transversality condition,
\beq
\label{transverse}
\p_\mu A_\mu^a = 0, 
\eeq
characteristic of the Landau gauge.  Property (ii) is the positivity of the Faddeev-Popov operator,
\beq
\label{psiMpsi}
(\psi, M(A) \psi) = (\p_\mu \psi, \p_\mu \psi) - (\psi^a f^{abc}, A_\mu^b \p_\mu \psi^c) \geq 0,
\eeq
for any wave function $\psi^a(x)$.  These two properties define the (first) Gribov region $\Omega$, and gauge fixing by this minimization produces configurations $A$ that all lie inside~$\Omega$.

\subsection{Basic idea of bound}

 It is very easy to establish bounds on configurations $A$ that are in the Gribov region $\Omega$.  Take any trial wave function $\psi(A)$ that may depend on $A$.  Then, from \eqref{psiMpsi}, it follows that every $A$ in $\Omega$ satisfies the bound,
\beq
\label{basicboundA}
(\psi^a(A) f^{abc}, A_\mu^b \p_\mu \psi^c(A)) \leq (\p_\mu \psi(A), \p_\mu \psi(A)).
\eeq
For an appropriately chosen trial wave function $\psi(A)$, an ellipsoidal bound on $A$ of the form,
\beq
\sum_k C_{k, \mu \nu}^{bc} a_{k, \mu}^{b*} a_{k, \nu}^c \leq  1,
\eeq
is obtained, as shown in Appendix A.  Here $a_{k, \mu}^b$ is the component of $A_\mu^b(x)$ in the fourier expansion,
\beq
\label{fouriercont}
A_\mu^b(x) = \sum_k a_{k,\mu}^b e^{ik \cdot x},
\eeq
on a finite periodic Euclidean volume, $V = L^d$, where $k_\mu = 2\pi n_\mu/L$, and $n_\mu$ runs over all integers.  Such a bound for a finite lattice was established in Appendix B of \cite{Zwanziger:1991}, and a stronger ellipsoidal bound is derived in the present article for continuum and lattice gauge fields in Appendices A and D respectively.  Upon taking expectation values, we obtain the bound on the gluon propagator $D_{\mu \nu}(k)$,
\beq
V^{-1} \sum_{k} \ C_{\mu \nu}^{bb}(k) D_{\mu \nu}(k) \leq  1,
\eeq
where we have used $\langle a_{k, \mu}^{b*} a_{k, \nu}^c \rangle =  V^{-1}\delta^{bc} D_{\mu \nu}(k)$.  In dimension 3+1, the continuum theory must be regularized, and in Appendix D, a bound on the lattice gluon propagator is derived from the positivity of the lattice Faddeev-Popov operator.  The limit in which the ultraviolet regulator is removed, $\Lambda \to \infty$, is discussed in sect. IV.

\subsection{Other gauge bosons}

The only input to the bounds obtained here is the restriction of the functional integral to the interior of the Gribov region.  For this reason, the bound is the same whether or not the gluons are coupled to quarks or not, although the bound becomes inconsistent in dimension 3+1 if the theory is not asymptotically free.  The bounds obtained here also apply to the propagator of other gauge bosons that belong to an $SU(N)$ gauge group, including those with Higgs coupling.  In the present article we are concerned with QCD gauge theory only.  However it should be noted that the Landau gauge is a special case of the $R_\xi$ gauge, with $\xi = 0$, that is used when the gauge field is coupled to a Higgs boson.  This gauge may be given a non-perturbative meaning by the minimizing gauge fixing described above.  This is straightforward in dimension 1+1 and 2+1, and our results hold in these dimensions.  In dimension 3+1, a lattice regularization of ultraviolet divergences would be required to give the theory a non-perturbative meaning, but we have not considered other gauge bosons in dimension 3 + 1.

\section{Bounds on gluon propagator}

\subsection{Bound on continuum gluon propagator}

  The continuum propagator is defined by
\beq
\label{defineprop}
\langle A_\mu^b(x) A_\nu^d(0) \rangle = V^{-1} \sum_k \delta^{bd} D_{\mu \nu}(k) e^{i k \cdot x},
\eeq
where the (hyper)cubic periodic volume $V = L^d$ is sufficiently large that (hyper)spherical symmetry holds, and the propagator has the tensor structure
\beq
D_{\mu \nu}(k) = D(k) \Big( \delta_{\mu \nu} - {k_\mu k_\nu \over k^2 } \Big),
\eeq
by transversality.  Note that a factor of the coupling constant $g_0$ has been absorbed into the gauge field $A$, so $D = g_0^2 D_0$, where $D_0$ is the unrenormalized, canonical propagator.  

{\it Statement:}  For gauge fixing to the interior of the Gribov region $\Omega$, as in the minimization procedure described above, the gluon propagator $D(k)$ in $SU(N)$ gauge theory satisfies the bound
\beq
\label{contbound}   
 J  \equiv {d-1 \over d} {1 \over (2 \pi )^d} \int d^dk \ {D(k) \over k^2 } \leq N.
\eeq
This holds for Landau gauge, in which case $d$ is the dimension of space-time, and for Coulomb gauge, in which case $d$ is the dimension of ordinary space and $D(k)$ is the instantaneous spatial gluon propagator.  (Because of renormalization, dimension 3+1 requires a special discussion which will be given shortly.)  This is proven in Appendices A and B.

Lest it be thought that this bound is trivial, note that for the free propagator in Landau gauge, $D(k) = 1/k^2$, the bound is violated because of an infrared divergence of the $k$ integration in dimension 1+1, 2+1, and 3+1, and by an ultraviolet divergence of the $k$ integration in dimension $3+1$.

If the angular integration is performed, the bound reads
\beq
\label{contbounds}   
J \equiv {d-1 \over d} {S_{d-1} \over (2 \pi )^d} \int_0^{\infty} d k \ k^{d-3} D(k) \leq N,
\eeq
where $S_1 = 2 \pi$, $S_2 = 4 \pi$, and $S_3 = 2 \pi^2$.  Note that $J$ has engineering dimension 0 in all dimensions $d$.

\subsection{Bound on lattice gluon propagator}

To discuss the case of dimension 3+1, we must regularize and renormalize.  The lattice provides a convenient regularization, and a lattice analog of the bound \eqref{contbound} holds,
\beq
\label{latticebound}
J \equiv{1 \over d (2 \pi)^d} \int_{- \pi}^{\pi} d^dk  { \sum_\mu \cos^2(k_\mu/2) \ D_{\mu \mu}(k)  \over 4 \sum_\lambda \sin^2(k_\lambda/2) } \leq N,
\eeq
as is shown in Appendix D.  (We have used the same symbols $J, D_{\mu \nu} ...$ for continuum quantities and their lattice analogs.)  Here we have taken the lattice volume to infinity keeping the lattice spacing finite, and the lattice momentum $k_\mu$ is a continuous angle $- \pi \leq k_\mu \leq \pi$.  The lattice propagator is given by $D_{\mu \nu}(k) =  \sum_x \langle A_{x, \mu}^b A_{0, \nu}^c \rangle e^{- i k \cdot [x + (e_\mu/2) - (e_\nu/2) ] }$, where the lattice variable $A_{x, \mu}^b$ is defined in \eqref{defineA}.

\subsection{Renormalized form of the continuum bound}

The lattice variable $A$ goes over in the continuum limit to
\beqa
A & \to & g_0(\Lambda) A_0
\nonumber  \\
& = & g_0(\Lambda) Z_3^{1/2}(\Lambda) A_r,
\eeqa
where $g_0 = g_0(\Lambda)$ is the bare coupling constant that depends on the ultraviolet cut-off $\Lambda$, $A_0$ is the unrenormalized or canonical continuum gauge field, and $A_r$ is the renormalized continuum gauge field.  Consequently the lattice gluon propagator is related to the unrenormalized and renormalized gluon propagators by
\beqa
\label{renormD}
D & \to & g_0^2(\Lambda) D_0
\nonumber  \\
& = & g_0^2(\Lambda) Z_3(\Lambda) D_r,
\eeqa
where $D_r(k)$ is the finite, renormalized, continuum propagator.  The lattice momentum goes over to the continuum momentum by $k \to a k$, where $a$ is the lattice spacing, so the lattice integral $\int_{- \pi}^\pi d^dk$ goes over to the continuum integral with a cut-off $\int_{|k| \leq C\Lambda} d^dk$, where $\Lambda = 1/a$, and $C$ is a constant of order 1.  Consequently, for large~$\Lambda$ the lattice bound \eqref{latticebound} goes over to the bound,
\beqa
\label{regularizedboundd}
J & \equiv &  {d-1 \over d} {S_{d-1} \over (2 \pi)^d} \int_0^{C\Lambda}  dk \ k^{d-3} D(k)
\\   \nonumber 
 & = &  g_0^2(\Lambda) Z_3(\Lambda) {d-1 \over d} {S_{d-1} \over (2 \pi)^d} \int_0^{C\Lambda}  dk \ k^{d-3} D_r(k) \leq N.
\eeqa
This is the renormalized form of the continuum bound \eqref{contbounds} that holds in dimension 3+1, where $d = 4$ for the Landau gauge and $d = 3$ for the Coulomb gauge.  We will find that the limit $\Lambda \to \infty$ is independent of $C$.

\section{Discussion}   

\subsection{Infrared bound in Landau Gauge}

We first discuss the Landau-gauge case.  The bound \eqref{contbounds} is more stringent in the infrared in lower dimensions because of the factor $k^{d-3}$ in the integrand.  Since the bound is finite and the integrand is positive, it follows that in dimension $d = 1+1$ the propagator $D(k)$ must vanish at $k = 0$,
\beq
\label{boundind2}
\lim_{k \to 0} D(k) = 0 \  \  {\rm for} \ d = 1+1.
\eeq
However in dimension 2+1, the bound \eqref{contbounds} is compatible with a finite value for $D(0)$.  Interestingly that is precisely the behavior that has been found in numerical studies of the  gluon propagator in Landau gauge in dimension $1+1$ and $2+1$, as discussed in the Introduction.  $D(k)$ may even be singular at $k = 0$ for $d = 2+1$, provided that the strength of the singularity remains less than $1/k$,
\beq
\label{boundind3}
\lim_{k \to 0} k D(k) = 0 \  \  {\rm for} \ d = 2+1.
\eeq 
This condition forbids the existence of gluons of mass zero for $d = 3$.  

Numerical studies also indicate that in dimension $d = 2+1$ the propagator in Landau gauge is suppressed in the infrared, although not as severely as in dimension $d = 1+1$, with $D(k)$ {\it decreasing} with $k$ as $k$ decreases to 0, but approaching a finite value, $D(0) > 0$.  There is no other explanation for this otherwise counter-intuitive decrease besides the proximity of the Gribov horizon in infrared directions.  The bound \eqref{contbounds} for $d = 2+1$ does not require such a decrease, but only that $D(k)$ not diverge as strongly as $1/k$.  Thus it appears that this bound by itself does not fully express the strength of the dynamical consequences of the cut-off of the functional integral at the Gribov horizon in dimension 2+1.

According to the lattice bound \eqref{latticebound}, the lattice Landau propagator $D_{\mu \nu}(k)$ in dimension 3+1 cannot have a singularity as strong as ${1 \over \sum_\mu\sin^2(k_\mu/2)}$ at $k = 0$ (where $\cos(k_\mu/2) = 1$),
\beq
\lim_{k \to 0} \sum_\mu \sin^2(k_\mu/2) \sum_\nu D_{\nu \nu}(k) = 0.
\eeq
Thus the bound on the lattice gluon propagator in Landau gauge in 4 Euclidean dimensions does not tolerate a $1/\sin^2(k/2)$ singularity, that is to say, a massless lattice gluon, for any finite value of the lattice spacing $a$ or, in other words, for any finite value of the cut-off $\Lambda = 1/a$.  Moreover the finiteness of the renormalized continuum bound  \eqref{regularizedboundd} for large but finite $\Lambda$ implies
\beq
\label{infraredin4}
\lim_{k \to 0} k^2 D_r(k) = 0, \  {\rm for} \ d = 3+1.
\eeq
According to numerical studies discussed in the Introduction, in dimension 3+1 there appears to be a kind of shoulder in $D(k)$ at low $k$, with a finite value of $D(0) > 0$.  This is entirely consistent with the infrared bound obtained here, but again, as in dimension 2+1, the gluon propagator $D(k)$ is apparently more strongly suppressed in the infrared than required by the bound we have obtained.  Thus it appears that also in dimension 3+1, the bound obtained here does not by itself fully express the strength of the dynamical consequences of the cut-off of the functional integral at the Gribov horizon.

\subsection{High-momentum bound in Landau gauge}

The coefficient $g_0^2(\Lambda) Z_3(\Lambda)$ which appears in the renormalzied continnum bound \eqref{regularizedboundd} is either zero or infinite as $\Lambda \to \infty$, so it might be thought that if $g_0^2(\Lambda) Z_3(\Lambda)$ is zero, then the bound \eqref{regularizedboundd} is trivially satisfied by any finite renormalized propagator $D_r$, and if $g_0^2(\Lambda) Z_3(\Lambda)$ is infinite, then the bound implies that the renormalized propagator $D_r(k)$ vanishes identically for all $k$, in which case the theory is inconsistent.  However before coming to this conclusion, we must also consider the $\Lambda$-dependence introduced by the ultraviolet cut-off of the integral at $k = C\Lambda$.

In dimension $d = 3+1$, the bound \eqref{regularizedboundd} reads
\beq
\label{regularizedbounda}
J \equiv g_0^2(\Lambda) Z_3(\Lambda) {3 \over 32 \pi^2} \int_0^{C\Lambda}  dk \ k  D_r(k) \leq N.
\eeq
We evaluate $J$ using the perturbative renormalization-group according to which, asymptotically at large $\Lambda$,
\beq
\label{gzeroZ3}
g_0^2(\Lambda) \approx {1 \over 2 b \ln \Lambda},
\eeq
where $b$ is the leading coefficient of the $\beta$-function $dg_0/d\ln \Lambda = - b g^3 + O(g^5)$.  It is gauge-independent and has the value \cite{ItzyksonZuber:1980}, p. 653,
\beq
\label{bequals}
b = {1 \over (4 \pi)^2 } \Big( {11 N \over 3} - {2 n_f \over 3} \Big),
\eeq
where $n_f$ is the number of quarks in the fundamental representation.  The dependence of $Z_3(\Lambda)$ on $\Lambda$ is found from the perturbative renormalization group.  We have
\beq
\label{Z3}
Z_3 = 1 + c g_r^2 \ln \Lambda + O(g_r^4)
\eeq 
where, in Landau gauge for SU(N) gauge theory \cite{ItzyksonZuber:1980}, p. 589,
\beq
c = {1 \over (4 \pi)^2 } \Big( {13 N \over 3} - {4 n_f \over 3} \Big),
\eeq
and we have added the quark contribution.  According to the renormalization group we have
\beqa
{d \ln Z_3 \over d \ln \Lambda}\Big|_{g_r} & = & c g_r^2 + O(g_r^4)
\nonumber \\
& = & c g_0^2 + O(g_0^4)
\nonumber \\
& = & {c \over 2b \ln \Lambda} + ... ,
\eeqa
which is solved by 
\beq
Z_3(\Lambda) = z_3 (\ln \Lambda)^p,
\eeq
where
\beqa
\label{valueofp}
p & = & {c \over 2b}
\nonumber \\
& = & {13 N - 4n_f \over 22 N - 4 n_f},
\eeqa
and $z_3$ is a finite constant of integration.  Thus the renormalized bound at large $\Lambda$ reads  
\beq
\label{regularizedboundb}
J \equiv {3 z_3 \over 64 \pi^2 b} \ln^{p-1}\Lambda \int_0^{C\Lambda}  dk \ k  D_r(k) \leq N.
\eeq
If $p > 1$, the coefficient of the integral, $\ln^{p-1} \Lambda$,  diverges, so the bound, $J \leq N$, requires that the renormalized propagator, $D_r(k)$, {\it vanish} for all $k$, in which case the theory is inconsistent.

To see if this happens, we note from \eqref{valueofp}
that
\beq
p < 1,
\eeq
provided that the denominator is positive, that is, provided that $11N > 2 n_f$.  This is the restriction on the number of quarks for the theory to be asymptotically free, $b > 0$.  Thus the inconsistency is avoided provided that the theory is asymptotically free, as we now assume.  

Because we are in the case $p < 1$, the coefficient of the integral in \eqref{regularizedboundb} vanishes in the limit $\Lambda \to \infty$,
\beq
\lim_{\Lambda \to \infty} \ln^{p-1} \Lambda = 0.
\eeq
There are three possibilities.  (i) If the integral in \eqref{regularizedboundb} is finite, the bound is trivially satisfied --- and vacuous.  This also happens if the integral diverges at large $\Lambda$, but too weakly to compensate the vanishing of the coefficient.  (ii) If the integral diverges sufficiently strongly that $J$ diverges for $\Lambda \to \infty$, then the theory is inconsistent.  (iii)  If the integral has a divergence that precisely compensates for the vanishing of the coefficient at large~$\Lambda$, then a finite bound results.  

To find out which possibility is realized, we first note that any finite contribution to the integral is annihilated by the coefficient, and we may write the bound as
\beq
\label{regularizedboundc}
J \equiv {3 z_3 \over 64 \pi^2 b} \ \ln^{p-1}\Lambda \int_\mu^{C\Lambda}  dk \ k  D_r(k) \leq N,
\eeq
where $\mu$ is an arbitrary mass.  Only the asymptotic  form of $D_r(k)$ at large $k$  concerns us.

According to the Callan-Symanzik equation, the corrections to scaling in the gluon propagator are logarithmic at high momentum, and the renormalized gluon propagator has the asymptotic behavior
\beq
D_r(k) \approx {r  \over k^2 \ln^p k },
\eeq
where the same power $p$ appears here as in $Z_3$, and $r$ is a finite constant that depends on the normalization condition.  Since any finite contribution to the integral gets annihilated by the vanishing of the coefficient, $\ln^{p-1}\Lambda$ for $\Lambda \to \infty$, we may extend this asymptotic expression down to finite $k$, and the quantity $J$ in \eqref{regularizedboundc} is thus, asymptotically at large $\Lambda$, given by
\beqa
\label{intlogp}
J & = & {3z_3 r \over 64 \pi^2 b} \ \ln^{p-1}\Lambda \int_\mu^{C\Lambda} {dk / k \over \ln^p k } 
\nonumber \\
& = & {3 z_3 r \over 64 \pi^2 b} \ \ln^{p-1}\Lambda\ \Big[ { \ln^{1-p} (C\Lambda) - \ln^{1-p} \mu \over 1 - p } \Big],
\eeqa
and, with $p < 1$, we obtain the limit
\beqa
\lim_{\Lambda \to \infty} J & = & {3z_3 r \over 64 \pi^2 b} \ {1 \over 1- p}
\nonumber  \\
& = & {3 z_3 r \over 32 \pi^2} \ {1 \over 2b - c} 
\nonumber \\
& = & {z_3 r \over 2N}.
\eeqa
The result is independent of $\mu$ and $C$, and finite. From  $\lim_{\Lambda \to \infty} J \leq N$ we obtain the non-trivial bound, 
\beq
\label{UVbond}
z_3 r \leq 2N^2.
\eeq

Four comments:  (i) The trivial and inconsistent possibilities are avoided because the divergence of the integral compensates for the vanishing of the coefficient in the limit $\Lambda \to \infty$.

(ii) The bound on the high-momentum limit of the gluon propagator is rather unexpected because the only input to this bound is the restriction of the functional integral to the interior of the Gribov region, and it was generally believed that this has dynamical consequences only in the infrared.  Note however that the high-momentum bound occurs only in dimension 3+1.

(iii)  The renormalized bound \eqref{regularizedbounda} may be expressed in terms of unrenormalized quantities
\beq
\label{regularizedbounde}
g_0^2(\Lambda)  {3 \over 32 \pi^2} \int_0^{C\Lambda}  dk k \ D_0(k) = {z_3 r \over 2N} \leq N.
\eeq
Since the left-hand side is constructed out of unrenormalized quantities, it is independent of any renormalization scheme.  On the other hand it is finite and independent of $\Lambda$ at large~$\Lambda$.  As such, it is a renormalization-group invariant.  Thus the finite quantity $z_3r$ for which we have just established the bound $z_3 r \leq 2N^2$, is in fact a renormalization-group invariant, although this was not apparent from the way $z_3$ and $r$ were introduced.

(iv)  The number $n_f$ of quark flavors has dropped out of the inequality \eqref{UVbond}.

\subsection{Infrared bound in Coulomb gauge}

The minimal Coulomb gauge is obtained by minimizing the Hilbert square norm of the space components $A_i^b({\bf x}, t)$ on each time-slice $t$,
\beq
F_{\bf A}(g, t) = || {^g}A_i(t) ||^2 = \int d^d x \ |{^g}A_i({\bf x}, t)|^2,
\eeq
where $i = 1, ... d$, and the dimension of space-time is $d+1$.  Consequently the equal-time propagator 
\beq
D({\bf k})( \delta_{ij} - k_i k_j/{\bf k}^2 ) \delta^{bc} = \int d^dx \ e^{-i k \cdot x} \langle A_i^b ({\bf x}, t) A_j^c({\bf 0}, t) \rangle,
\eeq
satisfies in space dimension $d$ the bounds we have derived in Landau gauge in space-time dimension $d$.  The expectation-value is independent of $t$ by time-translation invariance.  For orientation purposes we note that in zeroth-order perturbation theory the equal-time Coulomb-gauge propagator is given by
\beq
D^{(0)}({\bf k}) = \int { dk_0 \over 2\pi } {1 \over k_0^2 + {\bf k}^2} = {1 \over 2 |{\bf k}|}.
\eeq  

For space dimension $d = 2$ we obtain from \eqref{contbounds}, 
\beq
\lim_{{\bf k} \to 0} D({\bf k}) = 0, \ \ \ \ \ {\rm for} \ d = 2,
\eeq
so the equal-time propagator vanishes at ${\bf k} = 0$, $D({\bf 0}) = 0$.  This states that a gluon of zero momentum cannot be created by applying the field $A_i^b({\bf x}, t)$ to the vacuum, and thus the would-be physical gluons exit the spectrum.  From \eqref{regularizedboundd}, we obtain, as in  \eqref{infraredin4},  
\beq
\lim_{{\bf k} \to 0}  |{\bf k}| D_r({\bf k}) = 0, \ \ \ \ \ {\rm for} \ d = 3,
\eeq
where $D_r({\bf k})$ is the renormalized, equal-time, space-space Coulomb-gauge propagator.

\subsection{High-momentum bound in Coulomb gauge}

For space dimension $d = 3$, we must consider regularization and renormalization, as in the Landau-gauge.  Although renormalization in Coulomb  gauge has not been established to all orders, we suppose that it is renormalizable, and that the perturbative renormalization-group holds.  

We proceed exactly as in the Landau gauge case, but with different values for the constants.  For space dimension $d = 3$, eq.~\eqref{regularizedboundd} reads
\beq
J \equiv g_0^2(\Lambda) Z_3(\Lambda) {1 \over 3 \pi^2} \int_0^{C\Lambda}  dk \  D_{\rm r}(k) \leq N,
\eeq
where $g_0(\Lambda)$ is the unrenormalized coupling constant, and $Z_3(\Lambda)$ is the renormalization constant for the space components, $A_{i, 0} = Z_3^{1/2} A_{i, {\rm r}}$, in Coulomb gauge.  (The apace and time components of $A_\mu$ renormalize differently in Coulomb gauge.)

The renormalization constant $Z_3$ of the space components $A_i$ of the gauge field is given by
\eqref{Z3} where, by eq.~(B.37) of \cite{Zwanziger:1998}, with $Z_3 = Z_{\bf A}^2$, the coefficient $c$ has the value, 
\beq
c = {1 \over (4 \pi)^2} \Big( 2N - {4 n_f \over 3} \Big),
\eeq
for $SU(N)$, and we have added the quark contribution.  
As in the Landau gauge, we have
\beq
Z_3(\Lambda) = z_3 \ln^p \Lambda,
\eeq
where $p = c/2b$.  Here $c, r$ and $z_3$ have values appropriate to the Coulomb gauge, and we obtain
\beq
\label{coulcoef}
J = {1 \over 2 b \ln \Lambda}  z_3 \ln^p \Lambda {1 \over 3 \pi^2} \int_0^{C\Lambda}  dk \ D_r(k)  
\leq N,
\eeq
where the gauge-independent quantity $b$ is given in \eqref{bequals}.
As in Landau gauge, the theory is consistent only if $p < 1$.  We find   
\beq
\label{coulombp}
p = {c \over 2b} = {3 N - 2 n_f \over 11N - 2 n_f}, 
\eeq
and, as in Landau gauge, we have $p < 1$, provided that the denominator is positive, $11N > 2n_f$.  This is, again, the condition on the number of quarks for the theory to be asymptotically free, as we now assume.  As before, the coefficient of the integral in \eqref{coulcoef} vanishes for $\Lambda \to \infty$, and any finite contribution to the integral is annihilated in this limit.

According to the Callan-Symanzik equation, the renormalized equal-time propagator has logarithmic corrections asymptotically at large $\bf k$,
\beq
D_{\rm r}({\bf k}) \approx {r  \over 2 |{\bf k}| \ln^p |{\bf k}| },
\eeq
where $p$ is given in \eqref{coulombp}, and, as in \eqref{intlogp}, we have
\beq
J = {z_3 \over 6 \pi^2 b } \ \ln^{p-1} \Lambda \int_\mu^{C\Lambda}  dk \  {r \over 2 k \ln^p k} \leq N.
\eeq
As in Landau gauge, this gives, asymptotically at large~$\Lambda$,
\beqa
J & = & {z_3 r \over 12 \pi^2 b (1 - p) }
\nonumber  \\ 
& = & {z_3 r \over 6 \pi^2  (2b - c) }
\nonumber  \\
& = & { z_3 r \over 2N}.
\eeqa
We thus obtain
\beq
\label{UVcoulomb}
z_3 r \leq 2N^2,
\eeq
the same bound as in Landau gauge and, again, the number of quark flavors has dropped out.

As in Landau gauge, the quantity $z_3 r$ is a renormalization-group invariant.  In Coulomb gauge the bound on this quantity governs the high-momentum limit of the space-components of the  equal-time gluon propagator $D({\bf k})$, whereas in Landau gauge it governs the high-momentum limit of the Lorentz-invariant propagator $D(k)$.

\section{Concluding remarks}

We have obtained the continuum and lattice bounds~\eqref{contbounds} and~\eqref{latticebound} on the gluon propagator, and the bound \eqref{regularizedboundd} on the renormalized gluon propagator which hold in Landau gauge, where $d$ is the dimension of space-time, and in Coulomb gauge, where $d$ is the dimension of space and $D(k)$ is the instantaneous spatial gluon propagator.  In space-time dimension 2, 3 and 4, these bounds imply restrictions on the infrared behavior of the continuum gluon propagator in Landau and Coulomb gauge that are more severe in lower dimension, and in space-time dimension 4 there is, unexpectedly, a restriction on the high-momentum behavior of the continuum gluon propagator in Landau and Coulomb gauge.

It would be of interest to test the lattice and continuum bounds using numerical lattice data for the gluon propagator in 2, 3, and 4 space-time dimensions in Landau and in Coulomb gauge.  It is possible that the bounds are not close to being saturated.

\bigskip
	  
{\bf Acknowledgements}\\
The author is grateful for  stimulating conversations with Laurent Baulieu and correspondence with Atsushi Nakamura.

\appendix

\section{Ellipsoidal bound on continuum configurations}

We shall establish an ellipsoidal bound on continuum configurations $A$ that lie inside the Gribov region $\Omega$.  We consider the $SU(N)$ gauge group on a periodic Euclidean volume $V = L^d$.  To start, we substitute into inequality~\eqref{basicboundA} the trial wave function
\beq
\label{trialwavef}
\psi(A) = \psi_0 - \alpha { P_{q} \over (M_0 - p^2) } M_1(A) \psi_0,
\eeq
where $M = M_0 + M_1$ is the Faddeev-Popov operator, with $M_0 = - \p^2$, $M_1^{ac}(A) = - f^{abc} A_\mu^b \p_\mu$.  This wave function is inspired by first-order perturbation theory, according to which the first-order change in the zeroth-order wave function $\psi_0$ is given by a similar expression, but we shall of course obtain an exact bound.  The plane-wave state
\beq
\psi_0 = V^{-1/2} \exp(i p \cdot x) \ \eta,
\eeq
is an eigenvector of $M_0$, $M_0 \psi_0 = p^2 \psi_0$, $\eta$ is an x-independent, normalized color-vector, and $p_\mu$ is an allowed momentum vector on the periodic Euclidean volume $V = L^d$, $p_\mu = 2 \pi n_\mu /L$, where $n_\mu$ is an integer.  The operator $P_q$ is the projector defined by the kernel,
\beq
P_q(x, y) = V^{-1} \sum_{k; |k| \geq |q|} e^{i k \cdot (x-y)},
\eeq
which projects onto the direct sum of eigenspaces of $M_0$ belonging to all eigenvalues $k^2$ of $M_0$ that are greater than or equal to some fixed eigenvalue $q^2$.  We stipulate that the inequalities,
\beq
\label{momentaineq}
k^2 \geq q^2 > p^2 \geq (2 \pi / L)^2,
\eeq
are satisfied so the denominator in \eqref{trialwavef} is always positive, and that $k_\mu, p_\mu$, and $q_\mu$ are allowed momentum vectors on the periodic volume $V = L^d$.  The quantities $\alpha$, $p_\mu$ and $q_\mu$ are at our disposal, and $\alpha$ will be a variational parameter.    Note that $\psi_0$ is independent of $A$, and $M_1 = M_1(A)$ is linear in $A$, so the trial wave function $\psi = \psi(A)$ has a piece that is independent of $A$ and a piece that is linear in $A$.

With this wave function we have, by \eqref{basicboundA}, for all $A \in \Omega$, 
\beq
(\psi, M \psi) = I(\alpha) = X - 2\alpha Y + \alpha^2 Z \geq 0,
\eeq
where
\beqa
X & = & (\psi_0, (M_0 + M_1) \psi_0)
\nonumber  \\
Y & = & (\psi_0, M_1 { P_{q} \over (M_0 - p^2) } M_1 \psi_0)
\nonumber  \\
Z & = & (\psi_0, M_1 { P_{q} \over (M_0 - p^2) }
\nonumber  \\
&& \ \ \ \ \ \ \ \  \times (M_0 + M_1)  { P_{q} \over (M_0 - p^2) } M_1 \psi_0),
\eeqa
and we have simplified $Y$ using $P_{q} M_0 \psi_0 = 0$.  Positivity of $I(\alpha)$ for all $\alpha$ implies
\beq
X \geq 0;    \ \ \ \ \  \ \ \ \ Z \geq 0.
\eeq
Moreover $I(\alpha)$ has a minimum at $\alpha = Y/Z$, from which we obtain the bound
\beq
Y^2 \leq XZ.
\eeq

Because $A$ appears only in $M_1(A)$ which is linear in $A$, there is a term in $Z$ which is cubic in $A$, whereas $X$ and $Y$ are at most quadratic in $A$.  We shall bound the cubic term by means of the following lemma.\\
{\it Lemma:}  For the $SU(N)$ group the bound
\beq
\label{absboundonM1}
(\omega, [ M_0 + M_1(A) ] \omega) \leq N^2 (\omega, M_0 \omega)
\eeq
holds for any $A$ in $\Omega$ and any wave function $\omega(A)$.\\
This lemma is derived in Appenix B of~\cite{Zwanziger:1991}, but we present the derivation here for completeness. \\
{\it Proof:}  Consider first the $SU(2)$ group.  We decompose $M_1$ into the sum of 3 operators, $M_1 = M_1^{(1)} + M_1^{(2)} + M_1^{(3)}$ that are each given by
\beq
\label{defineM1}
M_1^{(b)}(A)  = S^b H^b(A)
\eeq
(no sum on $b$), where $(S^b)_{ac} \equiv i \epsilon^{abc}$ is an angular momentum matrix in the spin-one representation that acts on color variables, and $H^b = i A_\mu^b \p_\mu$ is a Hermitian operator that acts only on space variables.  We first bound the operator $M_1^{(3)} = S^3 H^3$.  Let $\omega_\pm^a \equiv e_\pm^a \phi(x)$, where $\phi(x)$ is any function of $x$, and $e_\pm$ and $e_0$ are $x$-independent, normalized eigenvectors of $S_3$, $S_3 e_m = m e_m$.  We have
\beq
0 \leq (\omega_\pm, [M_0 + M_1(A)] \omega_\pm) = (\phi, M_0 \phi) \pm (\phi, H^3(A) \phi),
\eeq
where the inequality holds for all $A \in \Omega$, and we have used the fact that $(e_\pm, S^1 e_\pm) = (e_\pm, S^2 e_\pm) = 0$ because $S_1$ and $S_2$ are off-diagonal in the $S_3$ basis.  It follows that the inequality
\beq
|(\phi, H^3(A) \phi)| \leq (\phi, M_0 \phi) 
\eeq
holds for any $\phi(x)$ and all $A \in \Omega$.  We now decompose any wave function $\omega$ according to $\omega = e_+ \phi_+ +e_0 \phi_0 +e_- \phi_-$, and we have for all $\omega$ and all $A \in \Omega$
\beqa
|(\omega, M_1^{(3)}(A) \omega)| & = & |(\omega, S^3 H^3 \omega)|
\nonumber  \\
& = & |(\phi_+, H^3 \phi_+) - (\phi_-, H^3 \phi_-)|
\nonumber  \\
& \leq & |(\phi_+, H^3 \phi_+)| + |(\phi_-, H^3 \phi_-)|
\nonumber  \\
& \leq & (\phi_+, M_0 \phi_+) + (\phi_-, M_0 \phi_-)
\nonumber  \\
& \leq & (\omega, M_0 \omega).
\eeqa
The same inequality holds for  $M_1^{(1)}$ and $M_1^{(2)}$ which gives
\beq
|(\omega, M_1(A) \omega)| \leq 3 (\omega, M_0 \omega).
\eeq
For the $SU(N)$ group, the proof is identical except that there are $N^2 -1$ terms in $M_1$, which gives
\beq
|(\omega, M_1(A) \omega)| \leq (N^2 -1) (\omega, M_0 \omega),
\eeq
and \eqref{absboundonM1} follows $\Box$.   

We apply the lemma to Z which is of the form $Z = (\omega, M \omega)$, where
$\omega = { P_{q} \over (M_0 - p^2) } M_1 \psi_0$.  The lemma yields
\beq
Z \leq N^2 Z_0,
\eeq
where
\beq
Z_0 \equiv (\psi_0, M_1(A) { P_{q} M_0 \over (M_0 - p^2).^2 } M_1(A) \psi_0),
\eeq
and we obtain the bound.
\beq
Y^2 \leq N^2XZ_0.
\eeq
The gain here is that $Z_0$ is only quadratic in $A$, whereas $Z$ contains a term that is cubic in $A$.   

We further simplify the bound by comparing $Y$ and $Z_0$.  We insert a complete set of eigenstates,
\beq
\psi_{k, a} = V^{-1/2} e^{i k \cdot x} e_a,  
\eeq
where $e_a$ are a basis of color vectors, and obtain
\beqa
\label{YandZzero}
Y & = & \sum_{k, a; |k| \geq |q|} { 1 \over k^2 - p^2 } | (\psi_{k, a}, M_1 \psi_0) |^2
  \\
Z_0 & = & \sum_{k, a; |k| \ \geq |q|}  {k^2  \over (k^2 - p^2)^2 } | (\psi_{k, a}, M_1 \psi_0) |^2.
\eeqa
From the restriction $k^2 \geq q^2 > p^2$ it follows that
\beq
{k^2 \over k^2 - p^2 } \leq {q^2 \over q^2 - p^2 }
\eeq
and consequently
\beq
Z_0 \leq  {q^2 \over q^2 - p^2 } Y.
\eeq
This gives the bound
\beq
Z_0Y \leq {q^2 \over q^2 - p^2 } Y^2 \leq {N^2 q^2 \over q^2 - p^2 }XZ_0,
\eeq
and so
\beq
Y \leq {N^2 q^2 \over q^2 - p^2 } X.
\eeq

We next consider
\beqa
\label{evaluateX}
X & = & (\psi_0, (M_0 + M_1) \psi_0)
\nonumber  \\
& = & p^2 - i p_\mu  f^{abc} \eta^{a*}  a_{k = 0, \mu}^b\eta^c,
\eeqa
where we have used the fourier expansion \eqref{fouriercont}, and $a_{k = 0, \mu}^b$ is the 0-momentum componant of $A_\mu^b(x)$.  We next show that
\beq
\label{etabound}
\Big| f^{abc} \eta^{a*}  a_{k = 0, \mu}^b\eta^c \Big| \leq {2 \pi \over L },
\eeq
where $\mu$ is a fixed Lorentz index.  To do so, we use $(\omega, M(A) \omega) \geq 0$ for $A \in \Omega$, where $\omega = V^{-1/2} \exp(\pm i 2\pi x_\mu /L) \eta$.  We have
\beq
(\omega, M \omega) = \Big({2 \pi \over L} \Big)^2 \mp i \Big( {2\pi \over L} \Big) f^{abc} \eta^{a*}  a_{k = 0, \mu}^b\eta^c \geq 0,
\eeq
from which \eqref{etabound} follows.  This gives
\beq
\label{boundX}
X \leq p^2 +  (2\pi/L) \sum_\mu | p_\mu |,
\eeq
and we obtain the bound on Y,  
\beq
\label{quadraticfV}
Y \leq {N^2 q^2 \over q^2 - p^2 } \Big[ p^2 + (2\pi/L) \sum_\mu | p_\mu | \Big]. 
\eeq
Note that $Y$ is quadratic in $A$, while the right-hand side is independent of $A$, so this is an ellipsoidal bound on $A$, as advertised.

We next evaluate $Y$, eq.~\eqref{YandZzero}.  We have
\beqa
\label{kM1p}
(\psi_{k, a}, M_1 \psi_0) & = &  - i p_\mu f^{abc} V^{-1} \int d^dx \     A_\mu^b(x) e^{i (p - k)\cdot x}   \eta^c
\nonumber  \\
& = &  - i p_\mu  f^{abc}  a_{k - p, \mu}^b\eta^c.
\eeqa
This gives
\beqa
Y & = & \sum_{k, a; |k| \geq |q|}  {| p_\mu  f^{abc}  a_{k - p, \mu}^b\eta^c |^2 \over k^2 - p^2 }
\nonumber  \\
& = & \sum_{k, a; |k + p| \geq |q|}  {| p_\mu  f^{abc}  a_{k, \mu}^b\eta^c |^2 \over (k+p)^2 - p^2 },
\eeqa
and we have the bound
\beqa
\label{quadraticfVb}
f^{abc} f^{ade} \eta^{c*} \eta^e \ p_\mu p_\nu
 \sum_{k; |k + p| \geq |q|} { a_{k, \mu}^{b*} a_{k, \nu}^d  \over (k + p)^2 - p^2 } \ \ \ 
  \\     \nonumber
 \leq {N^2 q^2 \over q^2 - p^2 } \Big[ p^2 + (2\pi/L) \sum_\mu | p_\mu | \Big]. 
\eeqa
For each $p$ and $q$ satisfying $q^2 > p^2 \geq (2\pi/L)^2$, and for each color vector $\eta$, this is an ellipsoidal bound on the fourier componants $a_{k, \nu}^b$ that holds at finite Euclidean volume $V$ for all configurations $A \in \Omega$.  Geometrically speaking, the configurations $A$ that satisfy the bound \eqref{quadraticfVb} define an ellipsoid $E$ in configuration space (that depends on $p$ and $q$).  The Gribov region $\Omega$ is contained in $E$, and we have the inclusions
\beq
\Lambda \subset \Omega \subset E.
\eeq  
Here $\Lambda$ is the fundamental modular region which consists of the absolute minimum of the minimizing functional on each gauge orbit.

\section{Bound on continuum gluon propagator}

We convert the ellipsoidal bound on configurations, just obtained, to a bound on the gluon propagator by taking expectation values,
\beqa
\label{quadraticfVc}
f^{abc} f^{ade} \eta^{c*} \eta^e \ p_\mu p_\nu
 \sum_{k; |k + p| \geq |q|} { \langle a_{k, \mu}^{b*} a_{k, \nu}^d \rangle  \over (k + p)^2 - p^2 } \ \ \ \ \ \ \
  \\     \nonumber
 \leq {N^2 q^2 \over q^2 - p^2 } \Big[ p^2 + (2\pi/L) \sum_\mu | p_\mu | \Big]. 
\eeqa
From the fourier expansions, 
\beqa
\label{defineprop}
\langle A_\mu^b(x) A_\nu^d(0) \rangle & = & V^{-1} \sum_k \delta^{bd} D_{\mu \nu}(k) e^{i k \cdot x}
\nonumber  \\
& = & \sum_k \langle a_{k, \mu}^{b*} a_{k, \nu}^d \rangle e^{i k \cdot x},
\eeqa
where $D_{\mu \nu}(k)$ is the gluon propagator on the finite periodic volume $V$, we obtain
\beq
\langle a_{k, \mu}^{b*} a_{k, \nu}^d \rangle = V^{-1} \delta^{bd} D_{\mu \nu}(k),
\eeq
and the last bound becomes
\beqa
  V^{-1}\sum_{k; |k + p| \geq |q|} { p_\mu D_{\mu \nu}(k) p_\nu \over (k + p)^2 - p^2 }    \ \ \ \ \ \ \ \ \ \ \ \ \ \ \ \ \ \ \ \ \ \ \ 
  \nonumber  \\
\leq {N q^2 \over q^2 - p^2 } \Big[ p^2 +  (2\pi/L) \sum_\mu | p_\mu | \Big],
\eeqa
where we have used $f^{abc} f^{abe} = N \delta^{ce}$.
This is an exact bound on the gluon propagator on a finite periodic Euclidean volume $V = L^d$.

We take the infinite-volume limit, $L \to \infty$, keeping $k, q$, and $p$ finite, and obtain
\beq   
V^{-1}\sum_{k; |k + p| \geq |q|} { p_\mu D_{\mu \nu}(k) p_\nu \over (k + p)^2 - p^2 } \leq {N q^2 \over q^2 - p^2 } \ p^2.
\eeq
We divide out a factor of $p^2$,
\beq   
V^{-1}\sum_{k; |k + p| \geq |q|} {\widehat p_\mu D_{\mu \nu}(k) \widehat p_\nu \over (k + p)^2 - p^2 } \leq {N q^2 \over q^2 - p^2 },
\eeq
where $\widehat p_\mu = p_\mu / |p|$ is a unit Lorentz vector.  Recall that $|p|, \widehat p_\mu$ and $q_\mu$ are quantities at our disposal.  We take the limit $|p| \to 0$, keeping $\widehat p_\mu$ and $q$ fixed, which gives
\beq   
V^{-1}\sum_{k; |k| \geq |q|} {\widehat p_\mu D_{\mu \nu}(k) \widehat p_\nu \over k^2 } \leq N .
\eeq
We now take the limit $q \to 0$, and convert the sum to an integral, since we are in the infinite-volume limit, and obtain the bound
\beq   
{1 \over (2 \pi )^d} \int d^dk \ {\widehat p_\mu D_{\mu \nu}(k) \widehat p_\nu \over k^2 } \leq N.
\eeq
Spherical symmetry is regained in the infinite-volume limit so, by transversality, $D_{\mu \nu}(k) = D(k) ( \delta_{\mu \nu} - k_\mu k_\nu / k^2  )$, and the angular average over $k$ yields the bound
\beq
\label{contbounda}   
 {d-1 \over d} {1 \over (2 \pi )^d} \int d^dk \ {D(k) \over k^2 } \leq N.
\eeq

\section{Simple ellipsoidal bound on continuum configurations at infinite volume}

We note parenthetically that we could have taken the limit of large volume $V$, without taking expectation values.  In \eqref{quadraticfVb}, we take the limit $L \to \infty$ keeping $k$, $q$ and $p$ finite, divide by $p^2$, take the limit $p \to 0$, followed by the limit $q \to 0$, and we obtain the ellipsoidal bound
\beq
\label{quadraticfVa}
f^{abc} f^{ade} \eta^{c*} \eta^e \ \widehat p_\mu \widehat p_\nu
 \sum_{k} { a_{k, \mu}^{b*} a_{k, \nu}^d  \over k^2 }
 \leq N^2. 
\eeq
We obtain a simpler ellipsoidal bound by summing over a complete basis $\sum_{b = 1}^{N^2-1}\eta_b^{c*} \eta_b^e = \delta^{ce}$, and $\sum_{\lambda = 1}^d \widehat p_\mu^\lambda \widehat p_\nu^\lambda = \delta_{\mu \nu}$, which gives, 
\beq
 \sum_k { a_{k, \mu}^{b*} a_{k, \mu}^b   \over k^2 } \leq N(N^2 - 1) d.
\eeq

In position space this bound reads, by \eqref{fouriercont},
\beq
\int_V d^dx \  A_\mu^b(x)  [(- \p^2)^{-1} A]_\mu^b(x) \leq N(N^2 - 1) d V.
\eeq
For a configuration $A_\mu^b(x)$ that has compact support, this bound becomes vacuous in the limit $V \to \infty$ bcause the right-hand side diverges with $V$, while the left-hand side remains finite.  However for a typical gauge-fixed configuration (and here we may suppose that a lattice regularization is in place), the integral on the left is a bulk quantity of order $V$, and the bound is meaningful.  Bounds on lattice configurations are given in Appendix~D.

\section{Bound on lattice gluon propagator}

\subsection{Notation for lattice quantities}

   Lattice configurations are defined by link variables $U_{x, \mu} \in SU(N)$, that live on the link $\langle x, x + e_\mu \rangle$, where sites of the lattice are labeled (in lattice units) by integer $x_\nu$, and $e_\mu$ is a unit Lorentz vector in the positive $\mu$ direction.  Numerical gauge fixing is done by minimizing the function
\beq
F_U(g) =  \sum_{x, \mu} {\rm Re \ tr} (1 - {^g}U_{x, \mu}),
\eeq        
with respect to local gauge transformations $g_x$, where
${^g}U_{x, \mu} \equiv g_x^{-1} U_{x, \mu} g_{x + e_\mu}$ is the gauge transform of the configuration $U_{x, \mu}$ by $g_x$.  In practice there are many local minima, and the particular minimum chosen is algorithm dependent.  For our purposes, any minimum will do; the absolute minimum plays no special role.  The only properties we shall use are that at any local minimum (i) the functional $F_U(g)$ is stationary, and (ii) the matrix of its second derivatives is positive.  Property (i) gives the lattice transversality condition,
\beq
\label{transverselatt}
\sum_\mu (A_{x, \mu}^a - A_{x - e_\mu, \mu}^a) = 0, 
\eeq
where we have introduced the lattice variables,
\beq
\label{defineA}
A_{x, \mu}^b(U) \equiv - {\rm tr}[t^b(U_{x, \mu} - U_{x, \mu}^\dag)],
\eeq
that are a lattice analog of the continuum variables $A_\mu^b(x)$.  Property (ii) is the positivity of the matrix element
\beq
\label{FPoplatt}
(\psi, M(U) \psi)  \geq 0,
\eeq
for all $\psi_x^a$.  Here $M_{xy}^{ab}(U)$ is the lattice Faddeev-Popov matrix.  It is a real symmetric matrix that is conveniently expressed as
\beq
M = M_0 + M_1,
\eeq
where $M_0$ and $M_1$ are defined by the quadratic forms
\beqa
\label{defFPoplatt0}
(\psi, M_0 \psi)
 \equiv 
 \sum_{x, \mu} {\rm tr} [ \ - \ (\psi_{x + e_\mu}^* - \psi_x^*) \ \ \ \ \ \ \ \ \ \ \ \ \ \ \
 \nonumber \\
 \times (U_{x, \mu} + U_{x, \mu}^\dag) (\psi_{x + e_\mu} - \psi_x) \ ], 
\eeqa
where $\psi_x \equiv t^a \psi_x^a$ and $\psi_x^* \equiv t^a \psi_x^{a*}$, and
\beqa
\label{M1}
(\psi, M_1(A) \psi) & = & 
{\rm tr} \{ [ \psi_{x + e_\mu} , \psi_x] \ (U_{x, \mu} - U_{x, \mu}^\dag) \} 
\nonumber \\
& = & f^{abc} \psi_{x + e_\mu}^a A_{x, \mu}^b \psi_x^c
\nonumber  \\
& = & - (1/2) f^{abc} (\psi_{x + e_\mu} +\psi_x)^a
\nonumber  \\
&& \ \ \ \ \ \ \ \ \ \ \ \  \times
 A_{x, \mu}^b (\psi_{x + e_\mu} -\psi_x)^c. \ \ 
\eeqa
The relation to the continuum Faddeev-Popov operator~\eqref{psiMpsi} is apparent.  The last expression is real when $A$ satisfies the lattice transversality condition \eqref{transverselatt}.  Properties (i) and (ii) define the (first) lattice Gribov region~$\Omega$.

	We consider a hypercubic periodic lattice of volume $V = {\cal N}^d$,  where ${\cal N}$ is an integer and $x_\mu = 0, 1, ... {\cal N}-1, {\rm mod}({\cal N})$.  Fourier transformation is given by 
\beq
\label{fourier}
A_{x, \mu}^b = \sum_k a_{k,\mu}^b \exp[ik \cdot (x + e_\mu/2)]
\eeq
where $k_\mu = 2 \pi n_\mu /{\cal N}$, and $n_\mu = 0, 1... {\cal N}-1, {\rm mod}({\cal N})$.  The transversality condition \eqref{transverselatt} is diagonal in momentum space where it reads
\beq
\sum_\mu K_\mu a_{k, \mu}^b = 0,
\eeq
and we have introduced
\beq
\label{largeK}
K_\mu \equiv 2 \sin(k_\mu/2),
\eeq
and similarly $P_\mu \equiv 2 \sin(p_\mu/2)$, $Q_\mu \equiv 2 \sin(q_\mu/2)$.

\subsection{Ellipsoidal bound on lattice configurations}  

We proceed as in the continuum case.  The lattice Gribov region $\Omega$ is defined by the condition on (transverse) configurations $U$,
\beq
\label{firstlattineq}
 - (\psi, M_1(A) \psi) \leq (\psi, M_0(U) \psi),
\eeq
for all $\psi$, where $A = A(U)$.

As in the continuum case, the matrix $M_1(A)$ is linear in $A$.  However in the lattice case, $M_0(U)$ is not independent of the configuration $U$.  We nevertheless obtain a simple lattice bound by introducing the matrix~${\cal K}_0$ defined by
\beq
\label{defK0}
(\psi, {\cal K}_0 \psi)
 \equiv 
 \sum_{x, \mu} {\rm tr} [ \ - \ 2 (\psi_{x + e_\mu} - \psi_x)^*(\psi_{x + e_\mu} - \psi_x) 
 \ ],
\eeq
which is independent of $U$ \cite{Zwanziger:1991}.  The difference,
\beqa
(\psi, [{\cal K}_0 - M_0(U)] \psi)
 \equiv 
 \sum_{x, \mu} {\rm tr} [ \ - \ (\psi_{x + e_\mu}^* - \psi_x^*) \ 
 \nonumber \\
 \times (1- U_{x, \mu}) (1- U_{x, \mu}^\dag) (\psi_{x + e_\mu} - \psi_x) \ ], 
\eeqa
is manifestly positive for every lattice configuration $U$ and every trial wave function $\psi$, so we have 
\beq
(\psi, M_0(A) \psi ) \leq (\psi, {\cal K}_0 \psi)
\eeq
which, by \eqref{firstlattineq}, yields the inequality
\beq
- (\psi, M_1(A) \psi ) \leq (\psi, {\cal K}_0 \psi),
\eeq
for every configuration $U \in \Omega$ and every $\psi$.  Geometrically, it is natural to define a region $\Theta$ in configuration space by this condition,
\beq
\Theta \equiv \{U: - (\psi, M_1(A) \psi ) \leq (\psi, {\cal K}_0 \psi) \ {\rm for \ all} \ \psi \},
\eeq 
where $A = A(U)$ is transverse, and we have the inclusions
\beq
\Lambda \subset \Omega \subset \Theta.
\eeq
Here $\Lambda$ is the fundamental modular region which consists of every configuration that is the absolute minimum of the minimizing function on its gauge orbit.  We shall derive a bound on the lattice gluon propagator that holds for all transverse configurations in $\Theta$, which then holds {\it a fortiori} for all configurations in the Gribov region $\Omega$.  Because ${\cal K}_0$ is independent of $U$ and because $M_1(A)$ is linear in $A = A(U)$, the proof goes just as in the continuum case, but with $M_0 \to {\cal K}_0$.

   We define
\beq
\psi_0 = V^{-1/2} e^{i p \cdot x} \eta,
\eeq
where $V = {\cal N}^d$ and $p_\mu = 2 \pi n_\mu/{\cal N}$ is a lattice momentum, and we have
\beq
{\cal K}_0 \psi_0 = P^2 \psi_0,
\eeq
where $P_\mu$ is defined as in \eqref{largeK}.  The continuum proof goes through, with the substitutions $p^2 \to P^2, k^2 \to K^2, q^2 \to Q^2$, and $M_0 \to {\cal K}_0$, and \eqref{momentaineq} becomes
\beq
\label{momentaineqa}
K^2 \geq Q^2 > P^2 \geq 4 \sin^2( \pi / {\cal N}).
\eeq
By \eqref{M1},  equation \eqref{kM1p} gets replaced by
\beq
\label{kM1pa}
(\psi_{k, a}, M_1 \psi_0) =  - i f^{abc} \sum_\mu P_\mu \cos(k_\mu/2) \  a_{k - p, \mu}^b \ \eta^c,
\eeq
\eqref{evaluateX} gets replaced by
\beq
\label{evaluateXa}
X = P^2 - i \sum_\mu P_\mu \cos(p_\mu/2)  f^{abc} \eta^{a*}  a_{k = 0, \mu}^b\eta^c,
\eeq
\eqref{etabound} by
\beq
\Big| f^{abc} \eta^{a*}  a_{k = 0, \mu}^b\eta^c \Big| \leq 2 \tan(\pi / {\cal N}),
\eeq
\eqref{boundX} by
\beq
\label{boundXa}
X \leq P^2 + 2 \tan (\pi/{\cal N}) \sum_\mu | \sin p_\mu |,
\eeq
\eqref{quadraticfV} by
\beq
\label{quadraticfVa}
Y \leq {N^2 Q^2 \over Q^2 - P^2 } \Big[ P^2 + 2 \tan(\pi/{\cal N}) \sum_\mu | \sin p_\mu | \Big], 
\eeq
and finally \eqref{quadraticfVb} by
\beqa
\label{quadraticfVd}
f^{abc} f^{ade} \eta^{c*} \eta^e 
 \sum_k { C_\mu(k', p) a_{k, \mu}^{b*} \ a_{k, \nu}^d C_\nu(k', p)  \over K'^2 - P^2 } \ \ \ \ \ \ \ 
  \\     \nonumber
 \leq {N^2 Q^2 \over Q^2 - P^2 } \Big[ P^2 + 2 \tan(\pi/{\cal N}) \sum_\mu | \sin p_\mu | \Big],
\eeqa
where $k'_\mu \equiv (k + p)_\mu$, $K'^2 \equiv 4 \sum_\mu \sin^2(k'/2)$, $C_\mu(k', p) \equiv P_\mu \cos(k'_\mu/2)$, and the sum over $k$ is restricted by $K'^2 \geq Q^2$.
This is an ellipsoidal bound that holds on a finite lattice for every configuration $a_{k, \mu}^b$ in the lattice Gribov region $\Omega$.

\subsection{Bound on lattice gluon propagator}

Upon taking expectation values, we obtain a bound on the lattice gluon propagator,
\beqa
\label{latticeboundD} 
 V^{-1}\sum_k { C_\mu(k', p) D_{\mu \nu}(k) C_\nu(k', p)  \over K'^2 - P^2 } \ \ \ \ \ \ \  \ \ \ \ \ \ \ \  \ \ \ \ \ \ \ \ 
  \\     \nonumber
 \leq {N Q^2 \over Q^2 - P^2 } \Big[ P^2 + 2 \tan(\pi/{\cal N}) \sum_\mu | \sin p_\mu | \Big],
\eeqa
where $D_{\mu \nu}(k)$ is the gluon propagator on a finite lattice, and we have used
\beq
\langle a_{k, \mu}^{b*} \ a_{k, \nu}^d \rangle = V^{-1} \delta^{bd} D_{\mu \nu}(k).
\eeq
  
We now take the lattice volume to infinity, ${\cal N} \to \infty$, while keeping the lattice spacing finite.  The lattice momentum $k_\mu = 2 \pi n_\mu / {\cal N}$ becomes a continuous angle $ - \pi \leq k_\mu \leq \pi$.  We divide out $P^2$, take the limit $p \to 0$ and then $q \to 0$, and we obtain the bound on the gluon propagator on an infinite lattice,
\beq
\label{latticeboundDa} 
 \widehat P_\mu T_{\mu \nu} \widehat P_\nu
 \leq N,
\eeq
where $\widehat P$ is an arbitrary unit vector.  Here
\beq
T_{\mu \nu} \equiv {1 \over (2 \pi)^d} \int_{- \pi}^{\pi} d^dk  { \cos(k_\mu/2) \ D_{\mu \nu}(k) \cos(k_\nu/2)  \over 4 \sum_\lambda \sin^2(k_\lambda/2) }
\eeq
(no sum on $\mu$ or $\nu$) is a tensor that is invariant under the hypercubic symmetries.  It is thus of the form $T_{\mu \nu} = J \delta_{\mu \nu}$, where $J = T_{\mu \mu}/d$, and the bound on the lattice gluon propagator reads $J \leq N$, or 
\beq
{1 \over d (2 \pi)^d} \int_{- \pi}^{\pi} d^dk  { \sum_\mu \cos^2(k_\mu/2) \ D_{\mu \mu}(k)  \over 4 \sum_\lambda \sin^2(k_\lambda/2) } \leq N .
\eeq



\end{document}